# Three-dimensional photography by holography

Jose J. Lunazzi

Universidade Estadual de Campinas
Institute of Physics
C.P. 6165
13083-950 Campinas-SP, Brazil

**Abstract**
Color encoding of depth is shown to occur naturally in holograms that are reconstructed under white light illumination. It can be registered in a common color photograph, allowing a simple method of visual decoding by means of ordinary colored 3-D spectacles. The fundamental holographic equations and the photographic procedure required for maximum fidelity in three-dimensional reproduction are described. The result is a new kind of photograph that shows all of the views of the object in a continuous sequence. It permits an animated photographic representation and also makes it possible to adjust the degree of depth visualization when observed as a stereoscopic representation.

Subject terms: ho1ography; photography; stereoscopy; color encoding; three-dimensional imaging.

CONTENTS
1. Introduction
2. Spectral encoding of depth
3. Holophotography: storage of multiple views in a single photograph
4. Influence of the wavelength resolution
5. Experimental details
6. Conclusions
7. Acknowledgments
8. References

## 1. INTRODUCTION

Color encoding of various properties of objects by holograms has been demonstrated previous1y[1-3]. In this paper we describe how a hologram naturally encodes a three-dimensional image that, after being photographed, constitutes a perfect anaglyph to be observed through colored spectacles[*].

The technique employs the well-known effect of blurring by chromatic aberration[4] that appears whenever a hologram is made by following the classical procedures of Leith and Upatnieks[5] or Denisyuk[6]. When illuminated in white light, this hologram shows an interesting property, visible in the form of a spectral multicolor profiling of shapes in the scene, that we demonstrate to constitute the spectral encoding of depth. This effect is more apparent for those regions of the object that are more distant from the plane of the hologram. The angular displacement of an image point, corresponding to the change in reconstructing wavelength, permits a color photograph of the scene to register the depth of that point as a spectrum whose extension is proportional to the depth value. This horizontal widening of the spectral distribution of colors brings an effect equivalent to the superposition of parallax changes between many different (hypothetical) points of view in horizontal continuous distribution around the photographic camera position.

For one specific color we obtain the scene that corresponds to one point of view. A photograph of this effect then constitutes the superposition of a continuous sequence of views. So, by photographing the holographic scene in a single exposure, we obtain an anaglyphic photograph that can re-create the three-dimensional image of the original object when proper color filters are used for each eye. Selection of the two stereoscopic points of view is then accomplished by choosing the wavelength of transmission for the filters. What is performed in that way is the substitution of the color ability of the film for a depth-encoding property, giving images that can be reconstructed as a continuous sequence of many points of view, a property that was unique for holograms. Each view is encoded by a specific wavelength value, but even with the limitations of color reproduction by the three-chromatic procedure, we can observe this property of the photograph by illuminating it with a source whose wavelength can be tuned over the visible spectral range. The photograph in this case appears as a living image with changing perspective, giving the appearance of the cinematographic representation of an object that rotates around a vertical axis in the plane of the photograph. This new photographic property permits us to consider a kind of stereo photography that offers a continuous variation of points of view within the observation angle, for which we propose the term "holophotography."

---

[*]  J. J. Lunazzi, "Anaglyphic effects in the photography from white light holograms." available upon request to the author.

## 2. SPECTRAL ENCODING OF DEPTH

The lateral coordinate x' of the holographic image of a point a distance z from the holographic film (Fig. 1), assuming also plane waves at the incidence angle $\theta_R$ for the reference and reconstruction beams, can be expressed as[7]

$$x' = x + \frac{\lambda_R - \lambda}{\lambda} z \tan\theta_R \quad (1)$$

where x is the original lateral coordinate of the object and $\lambda_R$, and $\lambda$ are the reference and reconstruction wavelengths, respectively. By denoting $\Delta\lambda = \lambda_R - \lambda$, we can calculate the width $\Delta x$ of the pattern that corresponds to the illumination of the hologram by the two wavelengths $\lambda_R$ and $\lambda$ simultaneously:

$$\Delta x = x' - x = \frac{\Delta\lambda}{\lambda} z \tan\theta_R \quad . \quad (2)$$

This proportionality between $\Delta x$ and z indicates that points with differing depths result in images in two colors whose widths are proportional to their depths. When photographed from a distance not very close to the hologram plane, the color images of the object at the plane of the photographic film also have widths proportional to the original z coordinate.

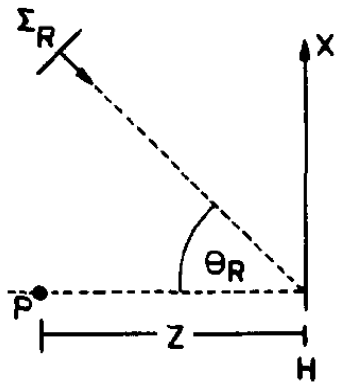

Fig. 1. Position of a point object P and reference (or reconstruction) wave $\Sigma_R$ with coordinate axis located at the hologram plane H.

Under white light reconstruction, the hologram of a point object yields an image in the shape of a line with continuously changing colors, constituting a spectral sequence. By the same reasoning, we see that for a line object in the vertical plane but inclined toward the plane of the holographic film, the resulting image in the final photograph will be a multicolored strip that widens for the points farther from the hologram plane.

Assuming now that the reference beam orientation corresponds to a direction of incidence located in a horizontal plane (from the left or the right side), it is not difficult to perceive that, in this way, depth can be easily color encoded in the color photograph of a hologram under white light reconstruction. This effect was conceived previously as blurring[4] or chromatic aberration. If takes one more step to demonstrate that the resulting image can be compared, even for extended objects, to a sequence of photographs obtained from many lateral points of view using colored filters in a spectral sequence when exposing the photograph.

We demonstrate this by means of Fig. 2. Figure 2(a) shows a double-exposed photograph registering two views of an object consisting of two points, O and A. The first position of the photographic camera is chosen, for convenience, as perfectly normal to a reference plane containing point O, with point A at a distance z behind it. For its second position the camera is moved laterally, keeping point O centered on its field of view and also maintaining a distance H from the camera to point O. The parameter i represents the constant distance between the camera lens and the image plane. By considering simple imaging properties we then obtain

$$\overline{A'A''} = \frac{iz \sin\theta_s}{H + z\cos\theta_s} \quad . \tag{3}$$

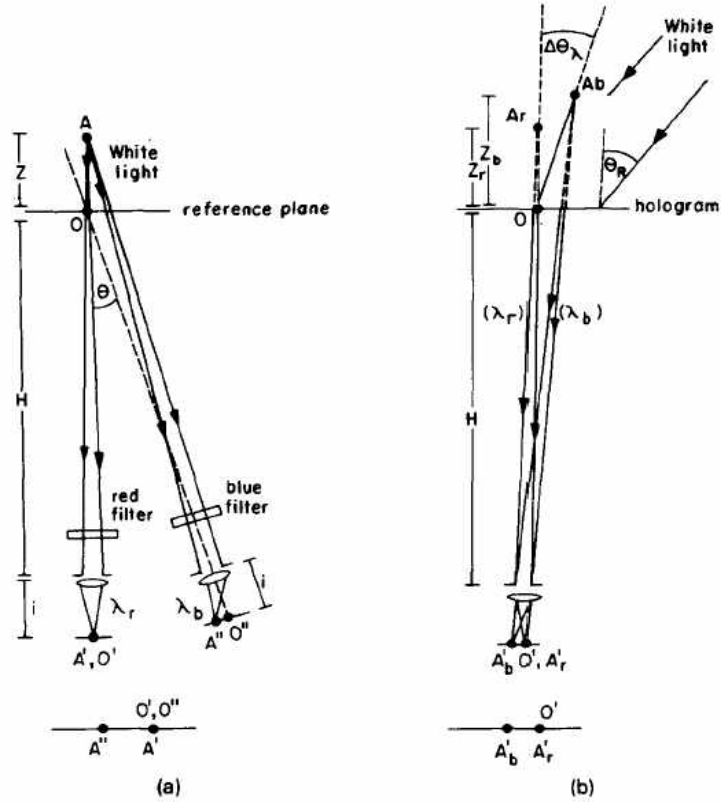

Fig. 2. Two cases of stereo photography: (a) conventional technique, in two views by double exposure and color filtering, and (b) holophotography, a single exposure with a hologram.

By the particular choice of points and camera positions, we obtain a result for the relative position of the image of a point in a two-exposure stereoscopic photograph, which we employ for any object point within the object field. The case of Fig.2(a) corresponds to a color photograph of a white object in white light but with a red filter interposed in front of the camera for the first exposure and a green filter for the second exposure. A stereoscopic representation can be obtained by observing this kind of photograph through ordinary (red and green) spectacles.

For the situation of photographing the hologram, Fig.2(b), corresponding to a single exposure of a holographic image employing ordinary color film, a reconstruction beam illuminates the hologram from the same direction as the original reference beam but in white light. Suppose the hologram was exposed in red laser light. Then point $A_r$ from the holographic image, corresponding to the red wavelength $\lambda_r$, coincides in the photographic image with the image of the reference point O in the hologram plane. Thus, image points $A'_r$ and O' coincide in the film plane. If z is the depth coordinate of the object point when the hologram was registered, a second point $A'_b$ is obtained in, for example, blue, from the wavelength $\lambda_b$ at a position defined by $z_b = z \lambda_r / \lambda_b$ at an angle $\Delta\theta_\lambda$. This angle can be by calculated by simple diffraction conditions or by using the equation for the holographic image of a point[7]:

$$\Delta\theta_\lambda = \arcsin[(1 - \frac{\lambda}{\lambda_R})\sin\theta_R] \quad . \tag{4}$$

We then obtain for the separation between image points $A'_r$ and $A'_b$ in the film the expression

$$\overline{A'_b A'_r} = \frac{i z_b \tan\Delta\theta_\lambda}{H + z_b} \quad , \tag{5}$$

from which we can compare the results of Figs. 2(a) and 2(b) by equating $M_s$, a parameter indicating stereoscopic magnification, defined as

$$M_s = \frac{\overline{A'_b A'_r}}{\overline{A'A''}} = \frac{\lambda_R (H + z\cos\theta_S)\tan\Delta\theta_\lambda}{\lambda(H + z\lambda_R/\lambda)\sin\theta_S} \quad . \tag{6}$$

$M_S = 1$ represents an optimum situation in which the photograph from a hologram is equivalent to a stereo anaglyphic pair since we are, for the moment, considering only two wavelengths from the white light beam. To obtain a result that is independent of the particular coordinate z of our object point, we must satisfy the condition

$$\theta_S = \arccos\frac{\lambda_R}{\lambda} \quad . \tag{7}$$

This yields $M_S = 1$ when the angle $\theta_R$ of the reference beam satisfies

$$\theta_R = \arcsin\{\frac{\sin\arctan[(\lambda/\lambda_R)^2 - 1]^{1/2}}{1 - \lambda/\lambda_R}\} \quad . \tag{8}$$

We considered many wavelengths in order to find an exact solution for conditions (7) and (8). The sign of $\Delta\theta_\lambda$ was not considered as fundamental because switching it is equivalent to interchanging the left and right eye filters, for example.

As an example of an exact solution, we can see in Table I one case that simultaneously satisfies both conditions and also gives the value $M_S = 1$. In the other cases of Table I, conditions (7) and (8) were not satisfied, and a value of $\theta_R$ was chosen such that the resulting angle $\theta_S$ is a consequence of the $\theta_R$ value and of condition $M_S = 1$. In the situations in which conditions (7) and (8) were not satisfied, we found that a value for $M_S$ very close to unity could be obtained, having only a slight dependence on the coordinate z. Two situations in Table I correspond to the practical, common case of exposing the hologram with light from a He-Ne laser. Table II shows how the angle $\theta_S$ was determined for that case, considering the extent of the object to be limited in depth to 50 mm in front or in back of the hologram and the distance H for the holographic camera as 200 mm. The values of the $\theta_S$ angles calculated in this way are smaller than the exact values that satisfy conditions (7) and (8) simultaneously. They can be calculated by the expression corresponding to the value $M_S = 1$ for the case z = 0:

$$\theta_S = \arcsin\frac{\lambda_R \Delta\theta_\lambda}{\lambda} \quad . \tag{9}$$

A direct consequence of this is that one can obtain a higher value by choosing a higher value for the reference angle $\theta_R$. In all cases, we see from Table I that the change in the stereoscopic magnification $M_S$ is not more than 15% for the typical situation we have considered, the same percentage corresponding to the change in the angle $\theta_S$, also shown in Table II as the extreme values giving $M_S = 1$ for z = 50 mm and z = -50 mm.

TABLE I. Selected examples of calculated values for the principal wavelength-dependent parameters. Wavelengths are in micrometers; angles are in degrees.

| $\lambda_R$ | $\lambda$ | $\theta_R$ | $\theta_S$ | $|\Delta\theta_\lambda|$ | $M_S$ |
|---|---|---|---|---|---|
| 0.42 | 0.78 | 79.5 | 57.5 | 57.5 | 1 |
| 0.42 | 0.69 | 45 | 17.8 | 26.7 | 1.07-0.90 |
| 0.42 | 0.68 | 45 | 17.3 | 25.6 | 1.07-0.90 |
| 0.55 | 0.63 | 45 | 5.1 | 5.8 | 1.03-0.96 |
| 0.45 | 0.65 | 45 | 13.1 | 18.1 | 1.06-0.91 |
| 0.63 | 0.5 | 75 | 14.5 | 11.5 | 0.94-1.11 |
| 0.63 | 0.5 | 45 | 10.7 | 8.4 | 0.95-1.10 |

TABLE II. Example of the identification of the proper angle $\theta_S$ for approximating the condition Ms = 1.

| $\theta_S[°]$ | $M_S(z=-50)$ | $M_S(z=0)$ | $M_S(z=50)$ |
|---|---|---|---|
| 9.8 | | | 1.00 |
| 10 | 1.14 | 1.04 | 0.98 |
| 10.5 | 1.08 | 0.99 | 0.94 |
| 11 | 1.04 | 0.94 | 0.89 |
| 11.4 | 1.00 | | |

The stereoscopic angle $\theta_S$ could appear, at first sight, as a false analog representation of the scene because the camera was in a single position when the scene was photographed. This is not so, however, since from Fig. 2(b) we see that different regions of the hologram are being used by the different wavelengths, sending many different views of the object to a single point of view for the observer of the hologram. In Fig.3 we see the same situation as in Fig. 2(b) but indicating the angle $\theta_H$ that represents the angular aperture from which the hologram is sending optical information of the object point A to the photographic camera. This angle can be compared to the stereoscopic angle $\theta_S$ by comparing triangles, thus obtaining the expression

$$\theta_H = \arctan\left[\frac{\lambda_R H \tan\Delta\theta_\lambda}{\lambda(H+z\lambda_R/\lambda)}\right] \quad . \tag{10}$$

which is not very different from the value corresponding to $\theta_S$ in Eq. (9), at least in our case of Table I, where the angle $\Delta\theta_\lambda = 8.4°$ allows us to make a paraxial approximation, and the variation due to the influence of the z value can also be considered of little significance.

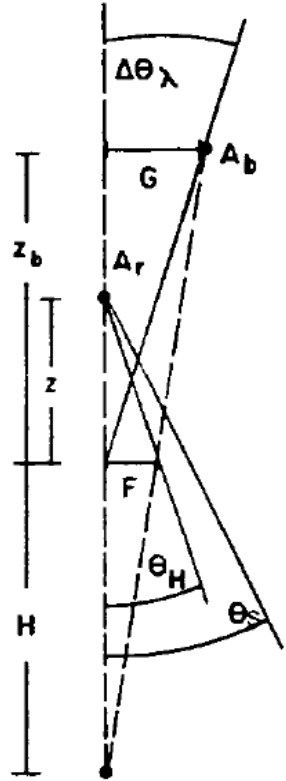

Fig.3. Comparison between the angle of view from the hologram $\theta_H$ and the stereoscopic angle $\theta_S$.

## 3. HOLOPHOTOGRAPHY: STORAGE OF MULTIPLE VIEWS IN A SINGLE PHOTOGRAPH

From the analysis of our technique we observe that the reconstruction in white light of the hologram being photographed corresponds to a situation in which the photographic film is exposed simultaneously to a continuous sequence of views, each in a slightly different wavelength. Although ordinary color photography reproduces color by using the registration of only three basic colors, the continuous sequence of views can be very well observed in the picture of a hologram. It is necessary only to illuminate the picture in white light and to interpose a variable color filter that can select, sequentially, the colors of the spectrum. An interference green filter can be tilted to transmit a continuous variation from green to red. The picture than becomes animated, showing the observer different views of the same object and giving the appearance of the object being rotated around an axis located vertically in the plane of the picture. It looks exactly like a movie in which the camera has revolved around the object. This new effect demonstrates the possibility of exchanging the capability of color reproduction for the capability of three-dimensional reproduction, being achieved in a continuous sequence of views. The sudden change in the color of the figure is not disturbing to the observer if we take care to prevent the same color-changing light from simultaneously illuminating any object in the surrounding area (the white frame of the picture, for example).

Another unique capability is that of watching a three-dimensional reproduction with the possibility of tuning its depth. By using a fixed filter for one eye and a color-variable filter for the other eye, we can observe the scene reproduced in variable depth, from zero to a maximum.

## 4. INFLUENCE OF THE WAVELENGTH RESOLUTION

For achieving perfectly resolved images, the continuous wavelength distribution that characterizes this technique of image reproduction requires perfect resolution of the wavelength at two stages: first, at the registering on color film, and second, when viewing through the eye filters. The presence of the continuous sequence of views differentiates the result from classical stereophotography, which involves two exposures, each with one color and view. Both techniques can be similar if only two monochromatic wavelengths are employed for reconstructing the hologram or if very narrowband filters (i.e., one red, one green) are interposed in sequence while the hologram is being photographed in two exposures.

We can formulate the general case by using a simplified model for the color response of a photographic film, considering a value $\Delta\lambda_F$ as representing the wavelength bandwidth of its response after being exposed to any monochromatic radiation of wavelength $\lambda$. The photographed point of coordinate x will then appear to have a horizontal extension:

$$\Delta x_F = \frac{\Delta\lambda_F z \tan\theta_R}{\lambda} \quad , \tag{11}$$

derived as a consequence of Eq. (2) since we are considering the common collection of a continuous sequence of points within a neighborhood defined by $\Delta\lambda_F$. For points or vertical lines having a dark background, the extreme colors of the spectral response must be seen on each side of the blurred image. For vertical borders in that situation, one of the two colors can be clearly seen because the spectral distributions do not overlap on any extreme position. Without considering any specific value for $\Delta\lambda_F$, we can just say that it constitutes a limiting factor, due to which condition Iz I < 50 mm has been considered in Tables I and II as an empirical limit.

The color response of a film for color reproduction allows us to predict that the range $\Delta\lambda_F$, cannot be independent of $\lambda$, but that it has minimum values at each wavelength that coincides with one of its three basic colors. Thus, a better sharpness can be obtained in the 3-D image if the bandwidths of the eye filters are made to coincide with those of the basic colors. It can also be easily understood that when the bandwidth of transmission $\Delta\lambda_E$ of the eye filter becomes larger than $\Delta\lambda_F$, the sharpness of the image will be reduced.

Another consideration is that the extension $\Delta\lambda_F$, when considered for the two eyes, does not double but remains nearly unaffected because of the mutual superposition of points that occurs in the binocular visualization[8]. One improvement that can be made is to have one eye filter as selective as possible and the other much more extended in bandwidth. The one permits reasonably good resolution while the other collects a more luminous image. The result is as if the visual system of the brain, when analyzing the three-dimensional scene, employs the sharpest representation for defining the lateral coordinates, and the more extended information is employed only for the determination of depth. It can also explain the great tolerance that exists for the difference in intensities between both images, since we have experienced observation with very different color responses, the red image sometimes being very weak, without affecting the results.

## 5. EXPERIMENTAL DETAILS

One of the first simple observations we made corresponded to the photograph of a white light hologram of a simple line object, a 3-D model of the capital letter E as printed in a color catalog[9]. We observed that all lines present spectral dispersion in one direction and no dispersion in a direction perpendicular to that one. Thus, the longer parts of the figure, which are oriented vertically in the natural position far reading the letter, are considered here as horizontals because it is necessary to rotate the figure by 90° to obtain the stereoscopic effect. These parts have a sharp representation, while parts of the figure in perpendicular positions showed an extended spectral distribution. From one extreme to the other of the blurred parts of the figure, it is observed that the dispersion of colors widens, continuously increasing and giving a figure like a prismatic rainbow. From Eq. (2), it can be interpreted that the spectral distribution widens for points that are farther from the hologram plane. Three long horizontal lines, for example, do not present dispersion. Since the upper one is the sharpest, it can be identified as coinciding, with the hologram plane. The other two lines appear more diffuse but have the same width and monochromaticity; they do not coincide with the hologram plane, and dispersion occurs along them. In some regions, both lines present dark bands (probably shadows) in which we can see the dispersion effect (some green and yellow colors). Those regions constitute horizontal discontinuities and every detail in the figure confirms our interpretation. On four of the remaining lines, nearly vertical ones, we see regions in which the dispersion is null but increases in the up and down directions. The orientations of the red and green lines change sides from the upper region to the lower region, corresponding to points that are in the back of the hologram plane, and when the line traverses that plane, it shows points that appear in front of the hologram plane. Using colored 3-D spectacles, we can see a clear 3-D image of the letter E.

Although the original hologram was of the reflection (Denisyuk) type, we can apply the same reasoning to it. The only particular effect we observed is that the points of no dispersion are represented in reddish color, probably due to the color selection generated by the Lippman-Bragg effect, other colors appearing only for points that are more deeply located. The characteristics of this chromatic selection by interference of white light seem to have very little effect on the resulting photograph since the two images in the stereo pair need not have exactly the same relative intensities. Also, in this case the effect of printing did not noticeably reduce the quality of the result.

In another experiment, the hologram was of an object made of fine wire, in a Denisyuk configuration with the reference beam incident at 45° from one side and the object very close to the holographic film at the other side. We used a common He-Ne laser and Agfa 8E75 film, processed with Kodak D-19 developer and Kodak R-10 bleaching bath. The image was observed in transmission, not in reflected light, and color pictures were exposed (15 min at f:22) in Kodak Ektachrome 100 ASA film by using a standard reflex camera with an f = 50 mm objective and one complementary lens (+3) for approximating to a distance of 20 cm to the hologram. Developed and enlarged to the original size of the object (9 cmX 12 cm) in a commercial standard color processing photographic laboratory, the result was a good-quality three-dimensional image of the object, observed by means of standard 3-D red-green spectacles. We can see it in Fig. 4, where the effect of the color spreading for line objects is very clear. For a few of them, one can observe that the yellow intermediary color is not present between red and green, as it show be. The only special care taken in photographing was to observe a position at which the red and green lines appearing on the borders of the holographic image could be considered as being of nearly equal intensities. The reference beam, for the scene to have its natural orientation, must be incident laterally and not from the top or bottom side of the object as is the usual case for white light holograms. We also photographed this hologram in two exposures by interposing in sequence the red and green filters from the spectacles, obtaining a picture in two colors in which some vertical lines could be seen clearly separated.

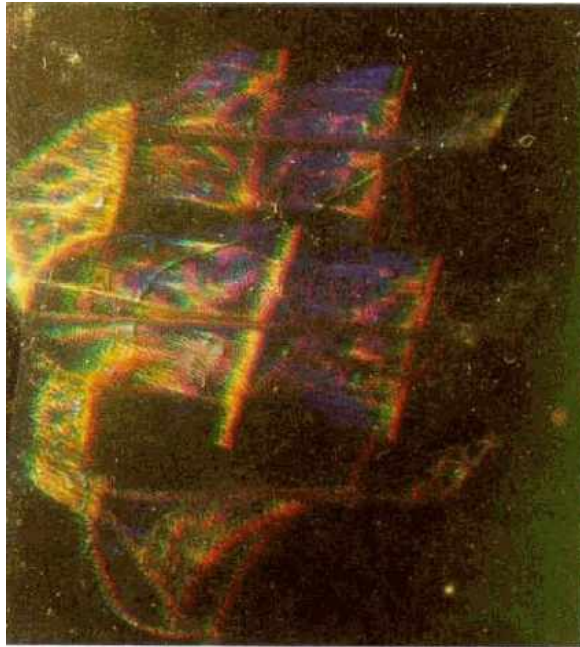
VIEW THROUGH ENCLOSED "3-D" GLASSES (RED-RIGHT EYE)
Fig. 4. Photograph of the ho1ogram of a wire object located a few millimeters behind the holographic plate.

As a third experiment, we phorographed a commercial hologram, "Rose", by Third Dimensions Ltd., of the reflection (Denisyuk) type. The image had a maximum depth of 50 mm in front or in back of the hologram and was photographed in the same way (60 s, f:32), resuhing in a 3-D image with more extension in depth than in the preceding cases. A certain blurring was observed corresponding to same parts at the bottom of the scene; nonetheless, the 3-D effect was impressive. In all of the photographs, and particularly those from the last case, the effect of animation was demonstrated by projecting the photographs through a slide projector and interposing a green interference filter tilted manually for the purpose of color selection. The rose appeared to rotate around an axis in the plane of the screen, and a dark cavity within the petals, for example, was stretched or reduced as a function of the color being selected. The degree of apparent rotation was not large, but with little effort, some small parts of the object could be seen to appear or disappear from behind others. Depth tuning was also achieved by using a red filter for one eye and, on the other eye, a green interferential filter that was tilted for the tuning.

As a final experiment, we performed with a He-Ne laser the hologram of a white metallic object representing our university's emblem, size 14 cm X 12 cm and 25 mm in depth. The size of this first hologram was 7 cmX 15 cm, and we projected its conjugate image together with a reference beam impinging the holographic film at 50°, making a second hologram 18 cm X 22 cm that was photographed under white light illumination at a distance of 60 cm in 100 ASA color film (2 s, f: I I) (cover photograph). Obliquity in the object position can be noticed because the first letters in the word UNICAMP appear in front of the plane of the photograph when observed with a red filter on the right eye and a green or blue filter on the left eye. The right-hand side of the emblem appears as penetrating in the plane of the photograph. Direct observation of the photograph allows one to identify the color spreading of the borders of the image according to their original position in depth, and the reversal of the spectral color sequence is very clear in the front and back side parts of the object.

## 6. CONCLUSIONS

A new technique for recording images allows storage of a continuous sequence of views from an object in color film, changing the color capability of the photosensitive material to a complete 3-D capability. Every perspective within a certain range is sent simultaneously to a single imaging system, being directly addressed by the parameter $\lambda$. This codification in wavelength has a simple and natural visual way of decoding by means of common colored 3-D spectacles. This new kind of photograph can also be shown in artificial animation by means of a simple color selector. The tuning of depth in the stereoscopic image is another interesting possibility. Although not perfect when performed with standard He-Ne lasers, an exact solution was demonstrated at extreme wavelengths of the visible region. The possibility of achieving a high degree of parallax was also demonstrated, a situation that could be experimentally performed in the future. The main element in the process is the effect of spectral dispersion, which has also been useful in the past for showing how some techniques of coherent optics can be similarly performed in incoherent light, employing much simpler equipment[10][11]. The result is also interesting as a new holographic property since it shows that the photograph of a hologram keeps the three-dimensional capability in a very natural way, putting holography in closer contact with photography as three-dimensional image registering techniques. We can even demonstrate how to obtain very similar results (holophotographs) by purely photographic means, which is the subject of our next paper[12].

## 7. ACKNOWLEDGMENTS

The author wishes acknowledge financial support by the Foundation of Assistance to Research – FAP of Campinas State University – Unicamp and to express his gratitude to Silvia Lunazzi for contributing creative ideas that stimulated this work. She, at the age of 9, had the idea of looking at photographs of holograms through colored 3-D glasses.

---

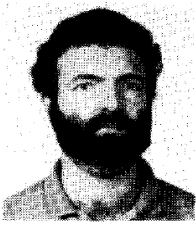José Joaquín Lunazzi was born In La Plata, Argentina, In 1948. He received his M.Sc. degree in 1970 and his Ph.D. degree in 1975, both in physics, from La Plata National University. He constructed his first hologram in 1969 as a consequence of his interest in stereo photography, which he performed at the age of 14. He joined Camplnas State University in 1976, working in optical metrology, Interferometry and holography. Dr. Lunazzi introduced holographic techniques in Latin America and has taught optics at every level, from postgraduate courses to courses for teachers at secondary and primary levels. He has also taught the use of simple materials in general courses for photographers, university students, artists, and the general public.

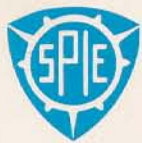

# Optical Engineering

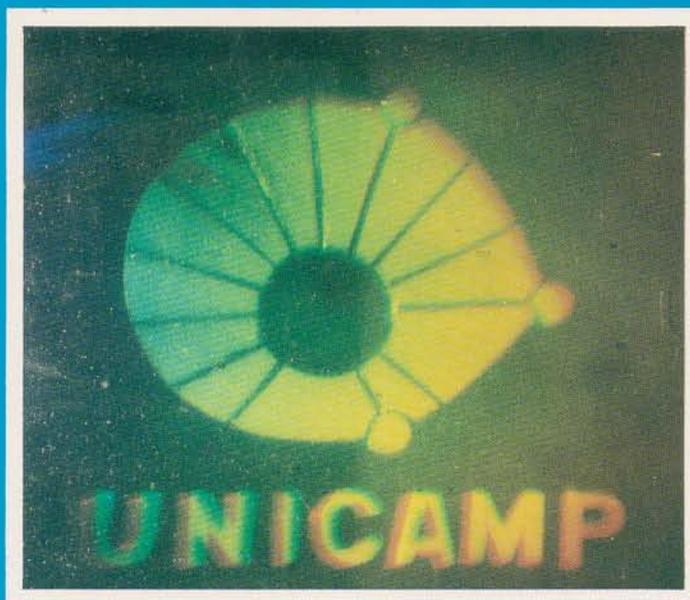

VIEW THROUGH ENCLOSED "3-D" GLASSES (RED—RIGHT EYE)

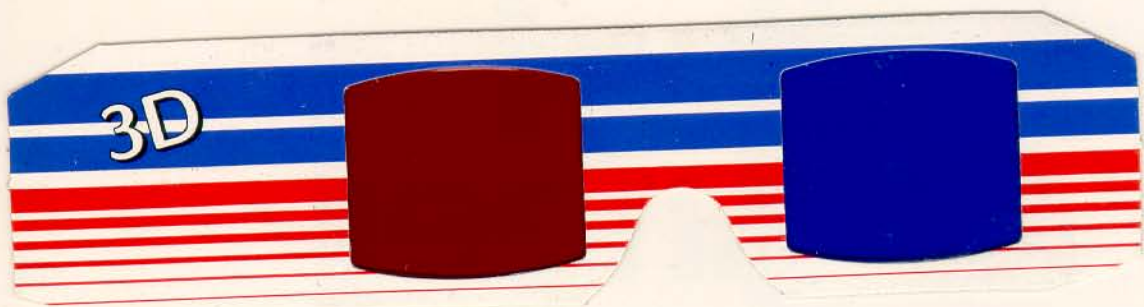

*Editorial*

Jack D. Gaskill, Editor

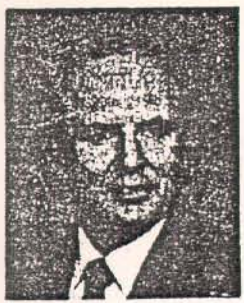

**A Hidden Three-Dimensional Effect**

The photograph on the cover of this issue of *Optical Engineering*, which is taken from Dr. José J. Lunazzi's paper entitled "Three-dimensional photography by holography," contains a hidden three-dimensional effect that can be observed by using the colored spectacles supplied with the issue. This effect is not produced by the usual red-green or blue-green composite photograph, called an anaglyph, but is a natural consequence of the process of recording a color photograph of an image formed by a white-light hologram. Following is a brief account of the events leading up to the publication of José's paper, in which an explanation of the effect is given.

It was over two years ago, while we were standing in the parking lot of the Town and Country Hotel in San Diego, that José introduced the subject of his paper to me. Although I had difficulty understanding the phenomenon he was patiently trying to explain to me, I became quite excited about it. In fact, I found it so fascinating that I immediately sought out two experts in the field of holography to get their opinions regarding the novelty and scientific merit of José's work. When these two experts responded with enthusiasm, I invited José to submit a paper to me for review and possible publication in the Journal, which he did. In addition, he submitted a second, related paper entitled "Holophotography with a diffraction grating," which also appears in this issue.

José first observed this phenomenon in 1984, and he credits his daughter, Silvia, with helping him discover it. He had just returned from an exhibit in Frankfurt, "Licht Blicke," with a catalog that contained over 100 color photographs of images formed by white-light holograms. As he was browsing through the catalog with Silvia, who was nine years old at the time, she suggested that they look at the pictures through some red-green 3-D spectacles that she happened to have. José admits that his first reaction was that of a typical adult—probably something like "oh, good grief"—but that he gave in and did as Silvia wished.

As they started looking at the pictures through the colored spectacles, José noticed that the color filtering produced some "nice" pseudo 3-D effects, but didn't think much about it—yet! Silvia soon tired of this activity, but José pressed on, systematically viewing image after image and occasionally observing an illusion of three-dimensional reality. Then, while viewing the 88th photograph, he suddenly became convinced that this 3-D effect was real—that an authentic representation of depth was present—and that he had discovered a stereo effect in the color photographs of holographic images. This led to several years of research, which yielded the two papers mentioned above.

It has been a lengthy, frustrating process getting these two papers published; José encountered delays in his research, there were postal strikes in Brazil, funding for the color photographs in the text and on the cover was difficult to identify, finding a supplier of the colored spectacles took time, etc. Everything eventually worked out, however, and we finally went to press. We hope that you will find this subject as fascinating and exciting as we did. In conclusion, I would like to thank José for his hard work and patience, Universidade Estadual de Campinas for its support of his research, and the National Science Foundation for providing partial funding for the printing of the color photographs.